\documentclass[apj]{emulateapj}

\usepackage{subfigure,captcont}

\usepackage{amsmath, amsfonts, amssymb, fancyhdr}
\usepackage{natbib}

\newcommand{\loh}{L_{\rm OH}}
\newcommand{\lfir}{L_{\rm FIR}}

\newcommand{\llfir}{\log(\lfir / L_\odot)}

\newcommand{\cmc}{{\rm cm}^{-3}}
\newcommand{\cms}{{\rm cm}^{-2}}
\newcommand{\kms}{{\rm km\; s}^{-1}}

\newcommand{\rh}{R_{\rm H}}
\newcommand{\rtwenty}{R_{\rm 1720}}
\newcommand{\rtwelve}{R_{\rm 1612}}

\begin{document}

\title{Constraints on OH Megamaser Excitation from a Survey of OH Satellite Lines}
\author{James McBride\altaffilmark{1}, Carl Heiles\altaffilmark{1}, Moshe Elitzur\altaffilmark{2}}

\altaffiltext{1}{Department of Astronomy, University of California, Berkeley, CA 94720-3411; jmcbride@astro.berkeley.edu}
\altaffiltext{2}{Department of Physics and Astronomy, University of Kentucky, Lexington, KY 40506-0055}
\begin{abstract}
We report the results of a full-Stokes survey of all four 18~cm OH lines in 77 OHMs using the Arecibo Observatory. This is the first survey of OHMs that included observations of the OH satellite lines; only 4 of the 77 OHMs have existing satellite line observations in the literature. In 5 sources, satellite line emission is detected, with 3 of the 5 sources re-detections of previously published sources. The 2 sources with new detections of satellite line emission are IRAS~F10173+0829, which was detected at 1720~MHz, and IRAS~F15107+0724, for which both the 1612~MHz and 1720~MHz lines were detected. In IRAS~F15107+0724, the satellite lines are partially conjugate, as 1720~MHz absorption and 1612~MHz emission have the same structure at some velocities within the source, along with additional broader 1612~MHz emission. This is the first observed example of conjugate satellite lines in an OHM. In the remaining sources, no satellite line emission is observed. The detections and upper limits are generally consistent with models of OHM emission in which all of the 18~cm OH lines have the same excitation temperature. There is no evidence for a significant population of strong satellite line emitters among OHMs. 
\end{abstract}
\keywords{masers --- galaxies: starburst --- galaxies: ISM --- radio lines: galaxies}

\section{Introduction} \label{intro}
OH megamasers (OHMs) are a class of luminous extragalactic masers that produce non-thermal emission in the 18~cm lines of the hydroxyl (OH) molecule, with two main lines at 1665/1667~MHz, and two satellite lines at 1612/1720~MHz. The names of OHMs are derived from their power relative to OH masers in star forming regions in the Milky Way---OHMs have typical isotropic luminosities of $10^3 L_\sun$, making them $\sim10^8$ times more luminous than Galactic OH masers. OHMs are primarily observed in galaxies experiencing a merger driven burst of star formation. Many were found in targeted searches of the luminous and ultra-luminous infrared galaxies [LIRGs have $\llfir > 11$ and ULIRGs have $\llfir > 12$, by definition] discovered by IRAS \citep{Darling2002a}.

OHMs are distinct from Galactic OH masers in star forming regions by more than just their luminosities. In particular, the 1665~MHz line is typically the brighter of the main lines in Galactic OH masers, while the 1667~MHz line is always brighter in OHMs. Galactic OH masers have narrow linewidths, typically narrower than 1~$\kms$, while individual components of OHMs have linewidths broader than 10~$\kms$, and total linewidths that measure $\sim$100--1000~$\kms$ (Lockett \& Elitzur 2008, hereafter LE08). There is diversity within the population of extragalactic masers, as not all share properties typical of OHMs. \citet{Henkel1990} (hereafter HW90) defined kilomasers as having isotropic luminosities $10^{-3} L_\odot < \loh < L_\odot$, and showed that kilomasers have properties distinct from those of megamasers. Kilomasers often feature a blend of absorption and maser emission, and show features not seen in megamaser sources. For instance, Very Large Array observations of the OH kilomaser in Messier~82 find some features with linewidths less than 10~$\kms$ and 1667/1665 ratios less than 1 \citep{Argo2007, Argo2010}.

The differences in masing OH line properties, from Galactic OH masers to powerful OHMs, reflect differences in the environment in which the masing occurs and in the mechanism by which the maser inversion is produced. Successfully modeling the properties of observed OHM emission thus provides constraints on the physical conditions of the galaxies in which OHMs are found. HW90, building on a model outlined in \citet{Henkel1987}, provided a simple explanation of the observed main line properties of OHMs in terms of low gain amplification of background radio continuum emission, and each of the lines having roughly equal excitation temperatures. The large linewidths of OHMs relative to Galactic OH masers can result in line overlap, as the thermal linewidth of FIR transitions between OH rotational levels can be larger than the energy separation between hyperfine levels within a single rotational level. When line overlap is important, equal line excitation temperatures is a natural result. \citet{Burdyuzha1990a} and \citet{Randell1995} also considered the excitation of OHMs. Though all three models came to slightly different conclusions about relative line strengths and the conditions in masing regions, all of the models supported radiative pumping of OHMs.

The first high resolution observations of the OHM Arp 220 by \citet{Lonsdale1994} found very compact maser structures. Two more OHMs, III~Zw~35 and IRAS~F17207--0014, were observed by \citet{Diamond1999} to have compact maser emission without corresponding compact background radio continuum features. These discoveries prompted consideration of whether collisional pumping played a role in producing compact components, as such small masing clouds within a much larger region of FIR emission seemed to require unrealistically high radiative pump efficiencies \citep{Lonsdale2002proc}. Observations of III~Zw~35 by \citet{Pihlstrom2001} found that the OH emission occurred in a ring, and concluded that radiative pumping and geometric effects could together explain the diffuse and compact maser emission. \citet{Parra2005a} performed more detailed modeling of III~Zw~35 that validated the results of \citet{Pihlstrom2001}, and they further suggested that such a model seemed to qualitatively explain other OHMs. The model parameters of \citet{Parra2005a} were then used by LE08 in their pumping analysis of OHMs. They successfully explained the main line ratios of OHMs and the weakness of satellite lines in the small number of OHMs in which they had been observed, and argued for radiative pumping via 53~$\mu$m emission being the dominant pumping mechanism in OHMs.

The relative ratios of the 18~cm OH lines are among the main observable parameters that models of the OHM environment and pumping should be able to explain. The sample of satellite line observations of extragalactic masers is, unfortunately, rather limited. \citet{Baan1987a} observed all four 18~cm lines in Arp~220, and found evidence for varying levels of excitation at different systemic velocities. \citet{Baan1989a} (hereafter BHH89) looked at the 1720~MHz line in 4 more OHMs, with detections in two: III~Zw~35 and IRAS~F17207--0014, and further noted a tentative detection of the 1612~MHz line in IRAS~20550+1655 with a peak flux density of 10~mJy, but did not provide a spectrum. \citet{Baan1992a} added detections of two more galaxies, Arp~299 at 1612~MHz and Mrk~231 at 1612~MHz and 1720~MHz. 

In their discussion of the satellite line results, \citet{Baan1992a} concluded that differences in optical depth could not fully explain the observed range of 18~cm line ratios, and suggested that this indicates a range of excitation temperatures in masing gas. In examining the existing 1720~MHz data, LE08 come to a different conclusion, saying that the handful of satellite line detections is consistent with the 1665~MHz, 1667~MHz, and 1720~MHz lines having roughly the same excitation temperature.

This work aims to address the question of whether any known OHMs display strong satellite line emission. The satellite line observations of OHMs prior to those presented here suggest that satellite lines should generally be quite weak relative to the main lines. Likewise, the models of OHM emission in HW90 or LE08 do not predict OHMs to have prominent satellite lines, given each of the 18~cm lines are expected to have roughly equal excitation temperatures. Discovery of a significant population of OHMs with satellite line emission in excess of that predicted for equal excitation temperatures would suggest that alternative pumping mechanisms play a prominent role in powering OHMs. Non-detections of satellite lines in the majority of OHMs, on the other hand, would be consistent with the expectations from the LE08 model.

\section{Sources}
The sources described here comprise the entire sample of extragalactic OH masers that are observable with the Arecibo telescope\footnote{The Arecibo Observatory is operated by SRI International under a cooperative agreement with the National Science Foundation (AST-1100968), and in alliance with Ana G. M\'{e}ndez-Universidad Metropolitana, and the Universities Space Research Association.} in Puerto Rico, which covers a declination range $-1^\circ \leq \delta \leq 38^\circ$. Two of these sources were recently discovered in a survey by \citet{Willett2012proc}, and the remainder of the sources were discovered by \citet{Darling2000,Darling2001,Darling2002a} or listed in the compilation of known OHMs in \citet{Darling2002a}. Of these sources, 4 had published satellite line observations prior to this survey, and were noted in the previous section. These are III~Zw~35, Arp~220, IRAS~F17207--0014, and IRAS~F20550+1655. To our knowledge, none of the remaining sources have published observations of either satellite line.

\section{Observations and Data Reduction}
The observations presented here used the $L$-band wide receiver on the 305~m Arecibo telescope, and occurred as part of a full-Stokes survey of OHMs from 2007 December--2009 December. The primary purpose of the survey was detection of Zeeman splitting in the strongest of the OH masing lines at 1667~MHz, for which results were presented in \citet{McBride2013}. The interim correlator at Arecibo can simultaneously perform full-Stokes observations of all four 18~cm OH lines, however, allowing the first comprehensive survey of satellite lines in OHMs as a ``bonus''. The boards of the interim correlator were configured such that 6.25~MHz bands were centered on the satellite lines at 1612~MHz and 1720~MHz. The other two boards observed the main lines; a 6.25~MHz board centered on the 1667~MHz line, and a 12.5~MHz board centered halfway between the 1665~MHz and 1667~MHz line. 

Most sources received a total of 3--4 hours of observing time, split equally on- and off-source. Position switching occurred every 4 minutes, with the off-source position located 4 minutes east of the source, to minimize the difference in hour angle between on- and off-source observations. The integration time was 1~s, to mitigate the effect of short duration RFI. 

As in \citet{Robishaw2008}, in which the detection of Zeeman splitting in OHMs was first demonstrated, and \citet{McBride2013}, we adopt the classical definition of Stokes~$I$, in which the flux is the sum of the two orthogonal polarizations rather than the mean. Thus our reported flux densities and integrated fluxes are a factor of two larger than some previously published results on OHMs, but we use self consistent definitions when making any comparison to previous results. More detailed discussion of the observations and data reduction methods used can be found in \citet{Robishaw2008} and in \citet{McBride2013}.

\subsection{Upper limits on non-detected lines}
For purposes of comparing detections and non-detections of lines that may appear in absorption or emission, we report velocity integrated line fluxes (Jy~$\kms$), rather than isotropic line luminosities. Line luminosities are useful for comparing the same line between different sources, while integrated line fluxes make comparisons of different lines in the same source more direct. Our method for defining upper limits on integrated line fluxes, denoted by $F$, for the 1612~MHz, 1665~MHz, and 1720~MHz lines is analogous to, but slightly less conservative than, that used in \citet{Darling2000, Darling2001, Darling2002a} with
\begin{equation}
    F = \sigma\; \Delta \nu_{1667}.
\end{equation}
Here, $\sigma$ represents the rms error in the spectrum at the expected location of the line. $\Delta \nu_{1667}$ is a measure of the width of the 1667~MHz line, defined as the frequency width in which 75\% of the line flux is contained. It is so defined to provide a more flexible measure of line width than the full width at half maximum (FWHM) for the complicated profiles often seen in OHMs, but gives the same answer for the case of a Gaussian profile. Effectively, this upper limit is a factor of 1.5 smaller than the \citet{Darling2000} definition, and allows incorporation of information from the 1667~MHz detection in setting an upper limit on satellite line emission. 

\section{Results}
For two sources, IRAS~F10173+0829 and IRAS~F15107+0724, we detected satellite 
line emission for the first time. Detailed discussion is provided for these
sources. We provide brief comments on detections of sources with existing
satellite line detections in the literature that were 
re-observed as part of the survey. We also discuss two sources with 
hyperfine ratios $\rh < 1.8$ over part or all of the spectrum, 
where $\rh$ is defined as the ratio of the integrated flux of the main lines, 
$F_{1667}/F_{1665}$. The remainder
of the sources observed in the survey, for which no satellite line emission
was detected, are not discussed in detail. Table \ref{tab:fluxes} 
lists the measurement of, or upper limits on, the integrated flux of 
each of the 18~cm OH lines for all sources detected at 1667~MHz. 
The non-detections or ambiguous detections, discussed in \citet{McBride2013},
are omitted. In some cases, it was not possible to provide a meaningful upper 
limit on non-detected lines. Omission of upper limits occurred in cases
of serious blending of the two main lines, in which case all flux was
attributed to the 1667~MHz line, or when Galactic HI or relatively time
stable RFI appeared at the expected location of the relevant 18~cm OH line.
\subsection{Notes on detections}
{\em IRAS F01417+1651 (III Zw 35).}
The 1720~MHz detection agrees reasonably well with that published in BHH89, so
we do not reproduce it here. RFI plagues the 1612~MHz spectrum, which 
prevents a detection, and results in relatively uninteresting upper limits.

\begin{figure}[h!]
    \centering
    \includegraphics[width=3.5in] {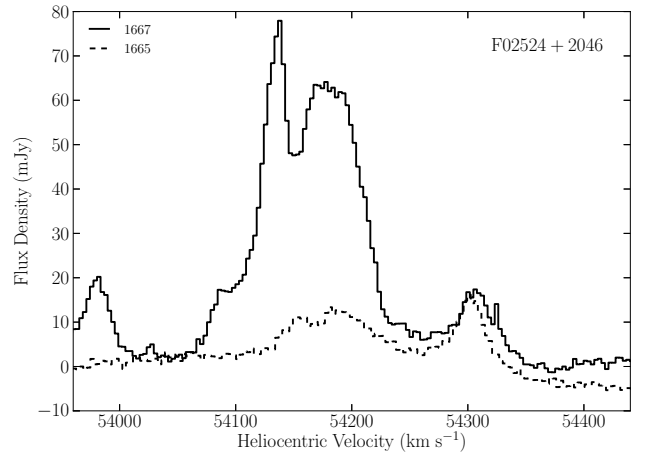}
    \caption{IRAS F02524+2046. The flux of the 1667 MHz line (solid) and the 1665~MHz line (dashed) are plotted on the same velocity axis for comparison of the two lines.} \label{fig:02524}
\end{figure}

\begin{table*}
    \centering
    \caption{OHM Line fluxes and limits} \label{tab:fluxes}
    \begin{tabular}{r r r r r r r}
            \hline \hline 
                   IRAS  & z & Integrated 1667~MHz & Integrated 1665~MHz & Integrated 1612~MHz & Integrated 1720~MHz & Notes \\
               FSC Name  &   & Flux (Jy km s$^{-1}$) & Flux (Jy km s$^{-1}$) & Flux (Jy km s$^{-1}$) & Flux (Jy km s$^{-1}$)  & \\ \hline 
            01417+1651   & 0.0274   & 53.8   & 7.2   & $<$1.3   & 0.5    & \\ 
            01562+2528   & 0.1658   & 4.9   & 0.7   & $<$1.1   & $<$0.8    & \\ 
            02524+2046   & 0.1814   & 9.6   & 4.0   & $<$0.7   & $<$0.6    & \\ 
            03521+0028   & 0.15206   & 1.0   & 0.12   & $<$0.7   & $<$0.5    & \\ 
            03566+1647   & 0.13352   & 0.8   & $<$0.3   & $<$40   & $<$0.7    & \\ 
            04121+0223   & 0.12183   & 0.7   & 0.2   & $<$0.4   & $<$1.9    & \\ 
            04332+0209   & 0.012014   & 0.9   & $<$0.12   & $<$0.15   & $<$0.2    & \\ 
            06487+2208   & 0.14334   & 2.8   & 0.4   & $<$0.4   & $<$0.3    & \\ 
            07163+0817   & 0.11097   & 0.6   & $<$0.2   & $<$0.3   & $<$3.0    & \\ 
            07572+0533   & 0.1898   & 0.7   & $<$0.12   & $<$0.4   & $<$0.2    & \\ 
            08071+0509   & 0.053463   & 3.1   & 0.6   & $<$1.9   & $<$0.3    & \\ 
            08201+2801   & 0.16769   & 8.0   & 0.9   & $<$0.9   & $<$0.8    & \\ 
            08279+0956   & 0.20864   & 1.8   & 1.0   & $<$1.1   & $<$1.4    & \\ 
            08474+1813   & 0.14541   & 1.4   & 0.2   & $<$1.3   & $<$1.4    & \\ 
            09039+0503   & 0.12514   & 1.9   & $<$0.3   & $<$0.4   & $<$10    & \\ 
            09531+1430   & 0.21486   & 3.0   & --   & $<$0.7   & $<$0.7    & (3) \\ 
            09539+0857   & 0.1289   & 12.2   & 3.8   & $<$0.9   & $<$1.7    & \\ 
            10035+2740   & 0.1662   & 1.1   & 0.14   & $<$1.0   & $<$0.9    & \\ 
            10173+0829   & 0.048   & 12.3   & 1.1   & $<$3.9   & 0.1    & \\ 
            10339+1548   & 0.19724   & 1.3   & 0.4   & $<$0.4   & $<$0.3    & \\ 
            10378+1108   & 0.1362   & 19.2   & --   & $<$1.3   & $<$4.3    & (1) \\ 
            11028+3130   & 0.199   & 1.9   & 0.4   & $<$0.2   & $<$0.7    & \\ 
            11180+1623   & 0.166   & 0.4   & $<$0.2   & $<$0.6   & $<$0.4    & \\ 
            11524+1058   & 0.18026   & 2.2   & 0.3   & $<$1.0   & $<$0.9    & \\ 
            12005+0009   & 0.1226   & 1.1   & 0.3   & $<$0.8   & $<$4.6    & \\ 
            12018+1941   & 0.16865   & 1.4   & $<$0.5   & $<$1.1   & $<$0.9    & \\ 
            12032+1707   & 0.21779   & 17.4   & --   & $<$6.4   & $<$2.1    & (1) \\ 
            12112+0305   & 0.073   & 12.9   & 1.5   & $<$0.9   & $<$3.9    & \\ 
            12162+1047   & 0.1465   & 1.0   & 0.06   & $<$0.8   & $<$0.8    & \\ 
           12243--0036   & 0.007048   & 1.4   & 0.5   & $<$1.3   & $<$0.5    & \\ 
            12549+2403   & 0.1317   & 0.7   & 0.06   & $<$0.5   & $<$0.3    & \\ 
            13126+2453   & 0.0112   & --0.7   & --0.6   & $<$0.3   & $<$0.2    & \\ 
            13218+0552   & 0.205   & 4.8   & --   & $<$2.7   & $<$2.5    & (1) \\ 
            14043+0624   & 0.1135   & 0.33   & 0.29   & --   & $<$0.8    & (2) \\ 
            14059+2000   & 0.1237   & 6.4   & 0.7   & $<$0.3   & $<$11.9    & \\ 
            14070+0525   & 0.265243   & 13.3   & --   & $<$4.4   & $<$2.4   & (1) \\ 
            14553+1245   & 0.1249   & 0.3   & $<$0.1   & $<$0.2   & $<$1.2    & \\ 
            14586+1432   & 0.1477   & 11.1   & --   & $<$2.1   & $<$2.0    & (1) \\ 
            15107+0724   & 0.012705   & 4.9   & 1.4   & 0.7   & --0.3    & \\ 
            15224+1033   & 0.1348   & 8.8   & --   & $<$52   & $<$2.4    & (1) \\ 
            15327+2340   & 0.018116   & 110   & 36.9   & $<$5.1   & 3.2    & \\ 
            15587+1609   & 0.13718   & 7.5   & 1.1   & $<$0.3   & $<$0.6    & \\ 
            16100+2527   & 0.131   & 0.8   & 0.4   & $<$15   & $<$0.3    & \\ 
            16255+2801   & 0.134   & 1.6   & $<$0.15   & $<$5.0   & $<$0.2    & \\ 
            16300+1558   & 0.24169   & 4.2   & --   & $<$2.4   & $<$2.4    & (2) \\ 
            17161+2006   & 0.1098   & 1.9   & 0.5   & $<$0.6   & $<$2.6    & \\ 
           17207--0014   & 0.0428   & 44   & 12.5   & $<$18   & 1.8    & \\ 
            17539+2935   & 0.1085   & 0.19   & $<$0.15   & $<$0.2   & $<$0.8    & \\ 
            18368+3549   & 0.11617   & 4.5   & 0.3   & $<$1.8   & $<$8.7    & \\ 
            18588+3517   & 0.10665   & 2.1   & 0.3   & $<$0.6   & $<$0.6    & \\ 
            20248+1734   & 0.129084   & 0.7   & --   & $<$1.1   & $<$1.8    & (2) \\ 
            20286+1846   & 0.134747   & 1.0   & 1.6   & --   & $<$0.5    & (2) \\ 
            20450+2140   & 0.12838   & 0.6   & $<$0.1   & $<$0.3   & $<$0.7    & \\ 
            20550+1655   & 0.036125   & 8.0   & --   & $<$0.7   & $<$0.5    & (2) \\ 
            21077+3358   & 0.176369   & 3.4   & 0.3   & $<$1.3   & $<$1.3    & \\ 
            21272+2514   & 0.150797   & 13.4   & 0.8   & $<$2.1   & $<$1.8    & \\ 
            22055+3024   & 0.126891   & 1.7   & 0.18   & $<$0.6   & $<$4.1    & \\ 
            22116+0437   & 0.19379   & 0.6   & $<$0.4   & $<$1.3   & $<$0.9    & \\ 
            22134+0043   & 0.212   & 4.7   & $<$0.25   & --   & $<$0.8    & (2) \\ 
            23019+3405   & 0.108   & 0.4   & $<$0.1   & $<$0.1   & $<$0.4    & \\ 
            23028+0725   & 0.1496   & 4.1   & 1.5   & $<$0.7   & $<$0.7    & \\ 
            23129+2548   & 0.17891   & 4.1   & 1.2   & $<$1.0   & $<$0.9    & \\ 
            23199+0123   & 0.1367   & 0.7   & $<$0.25   & $<$0.3   & $<$0.7    & \\ 
            23234+0946   & 0.1279   & 2.4   & --   & $<$0.5   & $<$4.6    & (1) \\ 
    \end{tabular}
    \tablecomments{
    (1) Emission is attributed to the 1667~MHz line because of blending;
    (2) Contaminant emission at expected location of line;
    (3) Conflicting evidence regarding the detection of the 1665~MHz line. See \citet{Darling2002a} for more discussion.
    }
\end{table*}
{\em IRAS F02524+2046.}
\citet{Darling2002a} detected this OHM, noting multiple matched components in the 1665~MHz and 1667~MHz lines. They provided hyperfine ratios for individual narrow features in the spectrum, with values $\rh = 1.4, 5.63, 1.88$ from high velocity to low. In our observations, shown in Figure \ref{fig:02524}, it is only possible to distinguish two peaks in the 1665~MHz emission that align with peaks in the 1667~MHz emission. The highest velocity feature is quite unusual, in that the 1665~MHz feature is so strong. The peak fluxes of the two lines are equal, and the ratio we find for the integrated flux at velocities 54,260--54,360~$\kms$ is $\rh = 1.3$, consistent with what \citet{Darling2002a} found. The hyperfine ratio is even smaller on the blue side of the line, as the 1665~MHz is moderately narrower than the 1667~MHz line and centered at a lower velocity. 

The observed hyperfine ratio in this velocity range is consistent with the range of 1--1.8 that is expected for thermal emission. To explore this possibility, we follow the example of \citet{Baan1982} in considering the required number of OH molecules to produce the observed line strength if it is optically thin thermal emission. Adjusting the equation they used for our definition of Stokes~$I$, the total number of OH molecules is
\begin{equation}
    N = 1.3 \times 10^{58} \alpha \left(\frac{D}{{\rm Mpc}}\right)^2 \left(\frac{F}{{\rm Jy\;}\kms}\right),
\end{equation}
where $\alpha$ takes values of 1 and 1.8 for 1667~MHz and 1665~MHz emission, respectively, $D$ is the distance in Mpc, and $F$ is the flux of the line integrated over velocity in units of Jy $\kms$. The required number of OH molecules is roughly $N = 10^{64}$ for each line, or a mass in OH of $\sim 1.5 \times 10^8 M_\sun$. For an OH abundance $n({\rm OH})/n({\rm H}_2)$ of order $10^{-6}$, this implies unrealistic quantities of molecular gas. Instead, the emission must include a significant non-thermal contribution, despite the hyperfine ratio that is consistent with thermal emission. 

The modeling of LE08 found that 1665~MHz maser emission was roughly as strong as1667~MHz emission for linewidths $\Delta V \simeq 2\;\kms$, and the 1665~MHz
line was stronger than the 1667~MHz line for narrower linewidths.
This was used to explain the observation that the 1667~MHz line is always 
stronger in OHMs, while the 1665~MHz line is typically stronger in Galactic 
OH masers, which have narrower lines. The total linewidth of the 
feature is considerably broader than that, and with this mechanism would
require many narrow features of comparable strengths that have blended 
together.

This OHM is notable in another respect, as it is among the most luminous known OHMs. 
The two most luminous OHMs, IRAS~F14070+0525 \citep{Baan1992b} and IRAS~F12032+1707 \citep{Darling2001}, both have such broad profiles that it is not possible to distinguish the two main lines. \citet{Baan1992b} nevertheless highlighted two pairs of features in the blended spectrum of IRAS~F14070+0525 that corresponded to the frequency separation of the two main lines, which would be consistent with strong 1665~MHz emission. The next most luminous OHM is IRAS~F20100--4156 \citep{Staveley-Smith1989}, for which only an upper limit on 1665~MHz emission could be placed. IRAS~F02524+2046 follows in this list of most luminous OHMs, and displays the unusual hyperfine ratio in part of its spectrum that we have discussed. The limited evidence regarding the most luminous OHMs is thus mixed, but hints at differences from the less luminous OHM population.  

\begin{figure}[h!]
    \centering
    \includegraphics[width=3.5in] {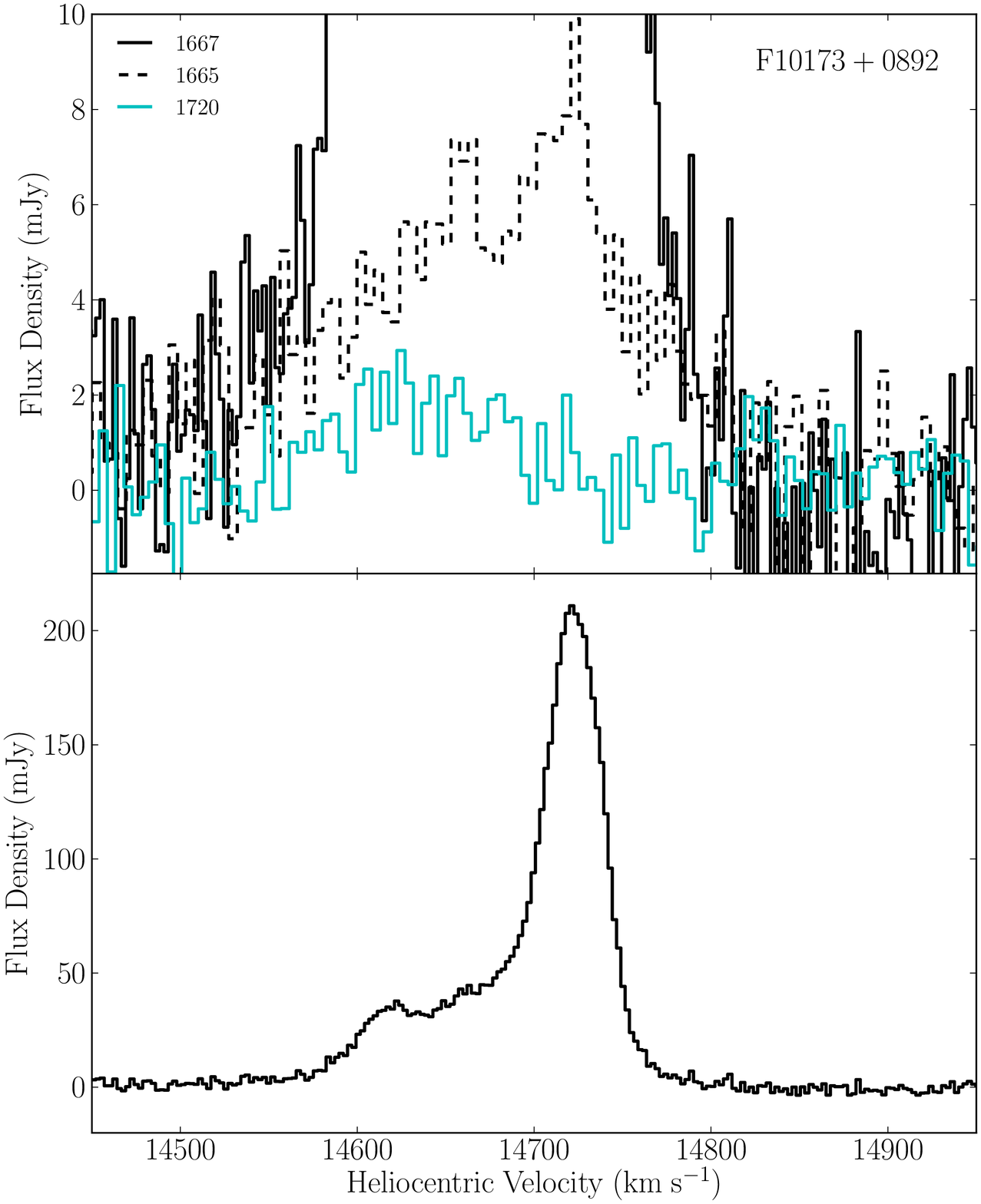}
    \caption{{\em Top panel:} The flux of the 1720~MHz line (solid cyan), the 1665~MHz line (dashed black), and the 1667~MHz line (solid black) are plotted on the same velocity axis. 
{\em Bottom panel:} The flux of the 1667~MHz line, shown on the same velocity axis, but with a flux density scale roughly a factor of 20 larger than that in the top panel, showing that the 1720~MHz emission occurs at the velocity of the weaker of the two 1667~MHz features.} \label{fig:10173}
\end{figure}

{\em IRAS F10173+0829.}
To the best of our knowledge, we present the first detection of 1720~MHz emission in this OHM. \citet{Baan1992a} noted that \citet{Mirabel1987} had previously looked for satellite lines in IRAS~F10173+0829, but discussion of this OHM in that paper is limited to the 1667/1665~MHz lines. Our detection of the 1720~MHz line is shown in Figure \ref{fig:10173}, along with the moderately stronger 1665~MHz line.
The 1720~MHz emission is produced at velocities 14,600--14,700~$\kms$, corresponding to the weaker of the two peaks in the 1667~MHz emission. There is negligible 1720~MHz emission, if any, at the same velocity as the stronger of the 1667~MHz peaks. The 1667:1665:1720 line ratio at the lower velocity peak is roughly 30:5:2. In this region, the 1720~MHz emission is considerably stronger than would be expected for equal excitation temperatures in the lines. In the higher velocity 1667~MHz peak, the hyperfine ratio is $R_H \sim 25$, and no 1720~MHz emission is visible. The limit on 1720~MHz in the higher velocity range does not rule out equal excitation temperatures of the lines in this region. When viewed over the entire range of velocities, the line ratios are consistent with equal excitation temperatures. 

{\em IRAS F14043+0624.}
\citet{Darling2002a} discovered this OHM and noted that it had an anomalous 
hyperfine ratio, with $\rh = 1.4$, while pointing out weak absorption at the 
edge of the 1665~MHz line. They suggested  
that a stable source of RFI could be present, and produce the anomalous
ratio. While we did see a strong, narrow spike of RFI near 1500~MHz, 
there was no evidence for RFI at the redshifted locations of the main lines 
between 1496--1499~MHz. RFI is often strongly 
polarized, but there is no structure in the Stokes $Q$, $U$, or $V$ spectra for 
IRAS~F14043+0624 at the frequency of the OH lines. Another line of evidence
disfavoring RFI as the cause of the feature is that none of the spectra of 
other sources at this frequency during our observations showed any evidence of
RFI. 
Nevertheless, the weak absorption that \citet{Darling2002a} observed at the edge of the 1665~MHz feature in their spectrum was not visible in our observations. 
We cannot rule out some form of RFI near the expected
location of the 1665~MHz line of this galaxy, even if none of our tests for
RFI within our data uncovered evidence for RFI.
\begin{figure}[h!]
    \centering
    \includegraphics[width=3.5in] {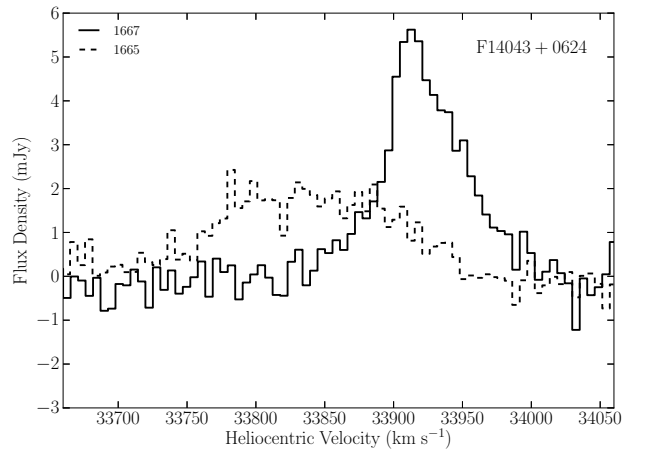}
    \caption{IRAS F14043+0624. The flux of the 1667 MHz line (solid) and the 1665~MHz line (dashed) are plotted on the same velocity axis for comparison of the two lines. } \label{fig:14043}
\end{figure}

Putting potential RFI issues aside, the hyperfine ratio we observed is $\rh = 1.1$, even smaller than that which \citet{Darling2002a} reported. However, given 
the line shapes are somewhat different, and the centers of the two lines are
offset by approximately 100~$\kms$, the hyperfine ratio may not actually be 
a meaningful measurement. 
The observations of the satellite lines do not help resolve the mystery, as
spectra for both lines are affected by RFI.

\begin{figure}
    \centering
    \includegraphics[width=3.5in] {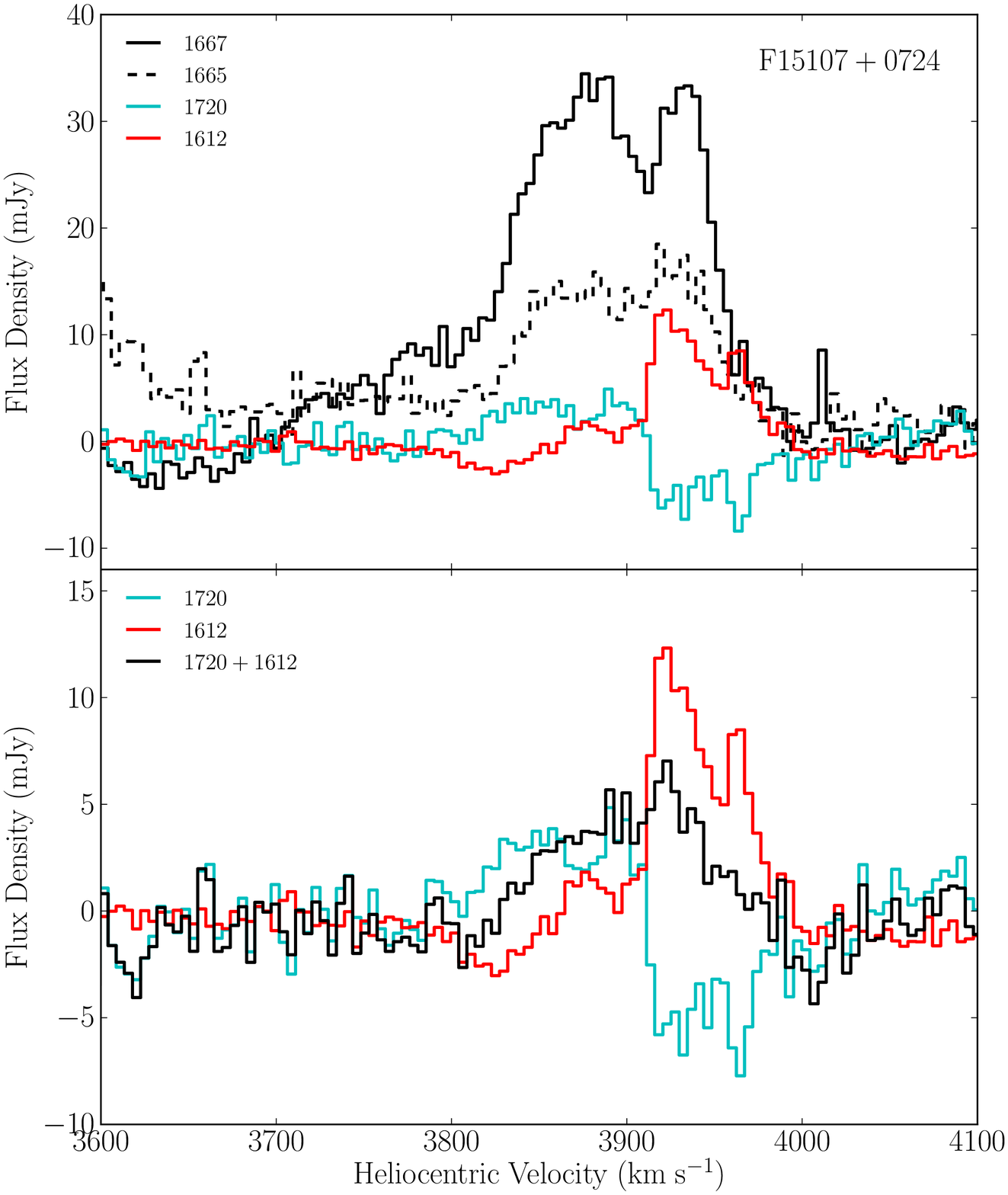}
    \caption{IRAS F15107+0724. {\em Top panel:} The flux of each of the 18~cm OH lines on the same velocity axes. {\em Bottom panel:} The flux of only the OH satellite lines, as well as the flux of the sum of the lines. While the two lines are not perfectly conjugate, much of the small scale structure in the satellite lines is conjugate, as evidence by the absence of small scale structure in the summed spectrum.} \label{fig:15107}
\end{figure}

{\em IRAS F15107+0724.}
This OHM was reported in \citep{Baan1987}, and is among the least intrinsically luminous OHMs, with an isotropic luminosity of $\sim 15\; L_\sun$. In our observations, all four 18~cm OH lines are detected (Figure \ref{fig:15107}). Emission in the 1612~MHz line falls predominantly between 3,910--3,980~$\kms$, and at its peak it is nearly as strong as the 1665~MHz line at the same velocity. Over this same velocity range, the 1720~MHz line appears in absorption, with a similar line shape, but at a weaker level than the 1612~MHz emission. The areas of conjugate line shape represent competition between the 1612~MHz and 1720~MHz transitions for pumping photons. 

\begin{figure}[h!]
    \centering
    \includegraphics[width=3.5in] {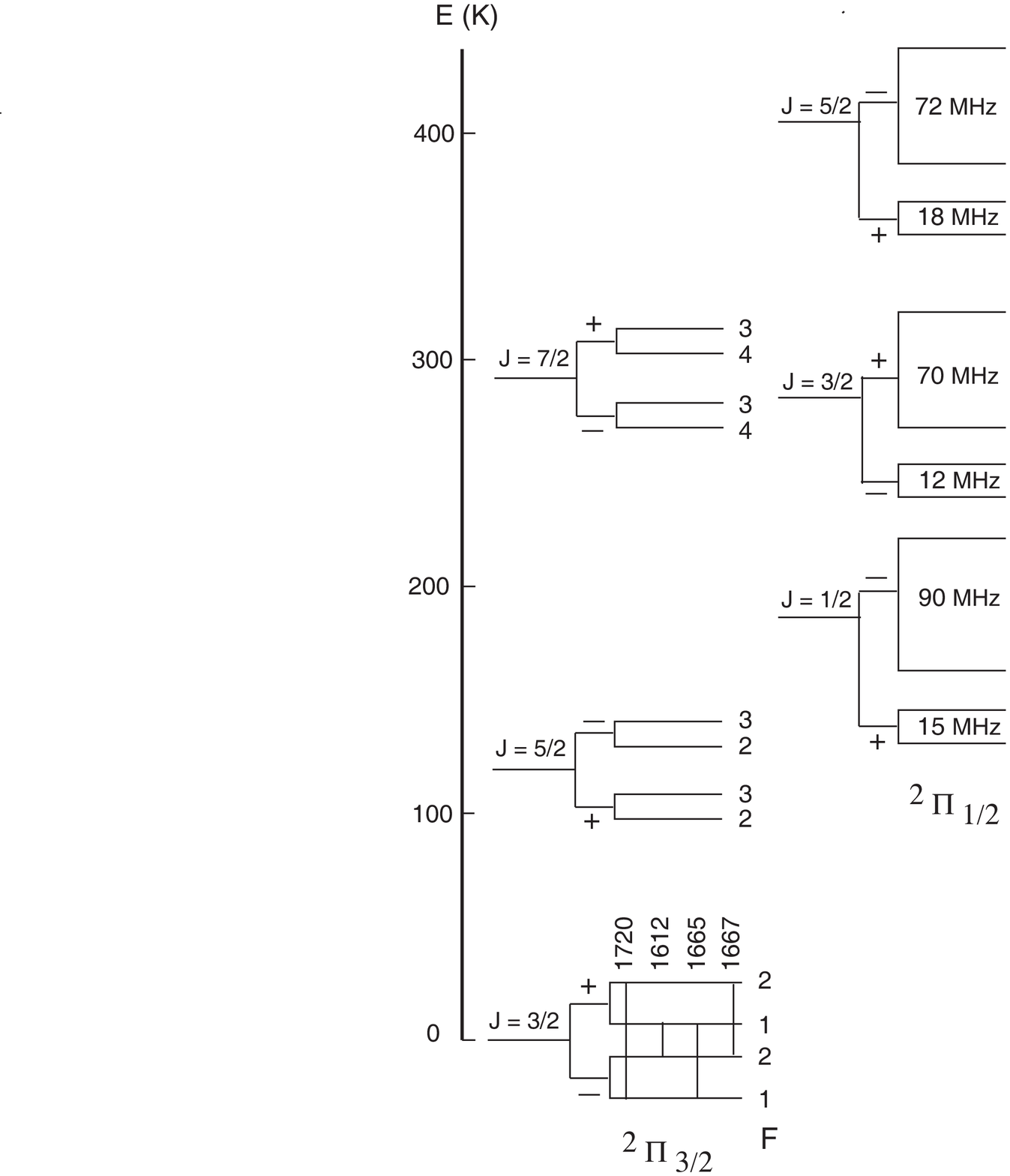}
    \caption{Rotational energy levels of OH, reproduced from \citet{Lockett2008}. There are four transitions within each rotational level as a result of $\Lambda$-doubling and hyperfine splitting (not shown to scale).} \label{fig:energy_level}
\end{figure}

The mechanism for producing conjugate emission is asymmetry in pumping that  
results from quantum mechanical selection rules in the OH rotational transition
ladder, shown in Figure \ref{fig:energy_level}. 
The 18~cm OH maser lines result from hyperfine transitions in the 
$^2\Pi_{3/2}(J=3/2)$ level, which is the ground state. The 1720~MHz line 
is produced by a transition from an $F=2$ to $F=1$ state, while 
the 1612~MHz line is a transition from $F=1$ to $F=2$ .
Transitions between rotational levels are
permitted when $|\Delta F| = 0, 1$. The $^2\Pi_{3/2}(J=5/2)$ (119~$\mu$ above 
the ground state) has hyperfine
levels with $F={2, 3}$, and thus will preferentially populate $F=2$ levels in
the ground state. The result is 1720~MHz inversion, and anti-inversion of 
the 1612~MHz line. A cascade from $^2\Pi_{1/2}(J=1/2)$ (79~$\mu$ above the 
ground state), which has hyperfine levels with $F={0,1}$, will overpopulate 
the $F=1$ levels in the OH ground state, producing 1612~MHz inversion and 
1720~MHz anti-inversion. 

For either of these pumping mechanisms, the FIR
transition must be optically thick. When both FIR lines are optically thick,
the 1612~MHz inversion dominates \citep{Elitzur1992book}. To produce 1720~MHz
inversion and conjugate 1612~MHz absorption requires the $^2\Pi_{5/2}(J=3/2)$
transition to be optically thick and $^2\Pi_{1/2}(J=1/2)$ transition 
to be optically thin. The result is that 
for a column density per velocity interval just below 
$10^{15}\; \cms\; {\rm km}^{-1}\;$s, 
the 1720~MHz line is inverted and the 1612~MHz line 
anti-inverted. The reverse behavior occurs just above 
$10^{15}\; \cms\; {\rm km}^{-1}\;$s \citep{VanLangevelde1995}. 

In the LE08 model, 53~$\mu$m radiative pumping of the main lines begins to 
occur at similar column densities per velocity interval at which
the conjugate satellite lines are produced. 
At velocities of 3900--4000 $\kms$, the ratio of line to continuum flux for
the 1667~MHz line gives an optical depth of --0.3, while the ratio of 
the 1667~MHz to 1665~MHz line gives an optical depth of --0.8. 
These are both consistent with weak amplification of the OH lines.  
While the conditions to produce the main lines and the conjugate satellite
lines lines appear similar, 
the line overlap needed to produce the inversion of the main lines
is not compatible with simultaneously producing conjugate satellite lines. 
Thus, within the same range of velocities in IRAS~F15107+0724, 
there must be two separate OH inversion mechanisms acting.

Conjugate OH satellite lines have been observed before in a diverse group of galaxies. Main line OH masing and absorption is seen in the starburst galaxies Messier 82 \citep{Seaquist1997} and NGC~253 \citep{Frayer2007}; in Centaurus A \citep{VanLangevelde1995}, the 18~cm OH main lines appear in absorption; in the distant radio galaxy PMN~J0134--0931 \citep{Kanekar2005}, the main lines also appear in absorption; and in another radio galaxy, PKS 1413+135 \citep{Darling2004, Kanekar2004}, the main lines were not detected. Our detection represents the first such example of conjugate emission in what otherwise appears to be a typical OHM, albeit one at the low end of the luminosity distribution. Both M~82 and NGC~253, which have lower FIR luminosities than IRAS~F15107+0724, are examples of kilomasers. HW90 argued that kilomasers represent a transition between powerful OHMs in LIRGs, and the OH absorbers found in more typical star forming galaxies. The observations of IRAS~F15107+0724 suggest that the transition from powerful OH masing to a mixture of masing and absorption may also feature increased satellite line strengths. The gain for each of the lines is low though, meaning moderately bright radio continuum emission is also required. 

{\em IRAS F15327+2340 (Arp 220).}
The 1612~MHz and 1720~MHz lines were first reported in BHH87, and the line ratios in different components and regions were discussed in detail. They concluded that there were differences in the excitation of the lines in each of the regions. Our 1612~MHz spectrum included features that could be astrophysical emission, but strong RFI produced serious structure in the bandpass that could not be removed. Parts of the line structure look quite similar to that published in BHH87, but given the superior quality of their spectrum, we do not provide our spectrum here. The 1720~MHz emission we see is consistent with that previously reported in BHH87.  

{\em IRAS F17207-0014.} Our re-detection of 1720~MHz emission agrees nicely with that published in BHH89, so it is not shown again here.

{\em IRAS F20550+1655.}
BHH89 noted a tentative detection of the 1612~MHz line with a strength 10~mJy at a velocity 55~$\kms$ above that of the 1667~MHz line. We confidently rule out a line of this strength, as our spectrum has an RMS error $\sim3$~mJy ($\sim6$~mJy in the ``classical'' definition), and no hint of a line is seen at the velocity given in BHH89.  

\subsection{Line excitation}
\begin{figure}
    \centering
    \subfigure[Excitation of the 1720~MHz line. Sources detected in each of the 1667~MHz, 1665~MHz, and 1720~MHz line in our survey are shown as black circles. The error bars reflect the error in the 1720~MHz flux, equal to the product of the line width and the rms error in the spectrum. Magenta squares are detections taken from the literature. Two types of arrows represent non-detections of one or more of the lines. Arrows pointing up represent sources in which the 1720~MHz line was not detected, but the 1665~MHz line was detected. Arrows pointed up and to the right are sources in which neither the 1665~MHz nor the 1720~MHz line was detected. The solid line shows the expected flux ratios for equal excitation temperatures of the lines when the optical depth of the 1667~MHz is varied, given by Equations \ref{eq:excitation1665} and \ref{eq:excitation1720}.]{
    \includegraphics[width=3.5in] {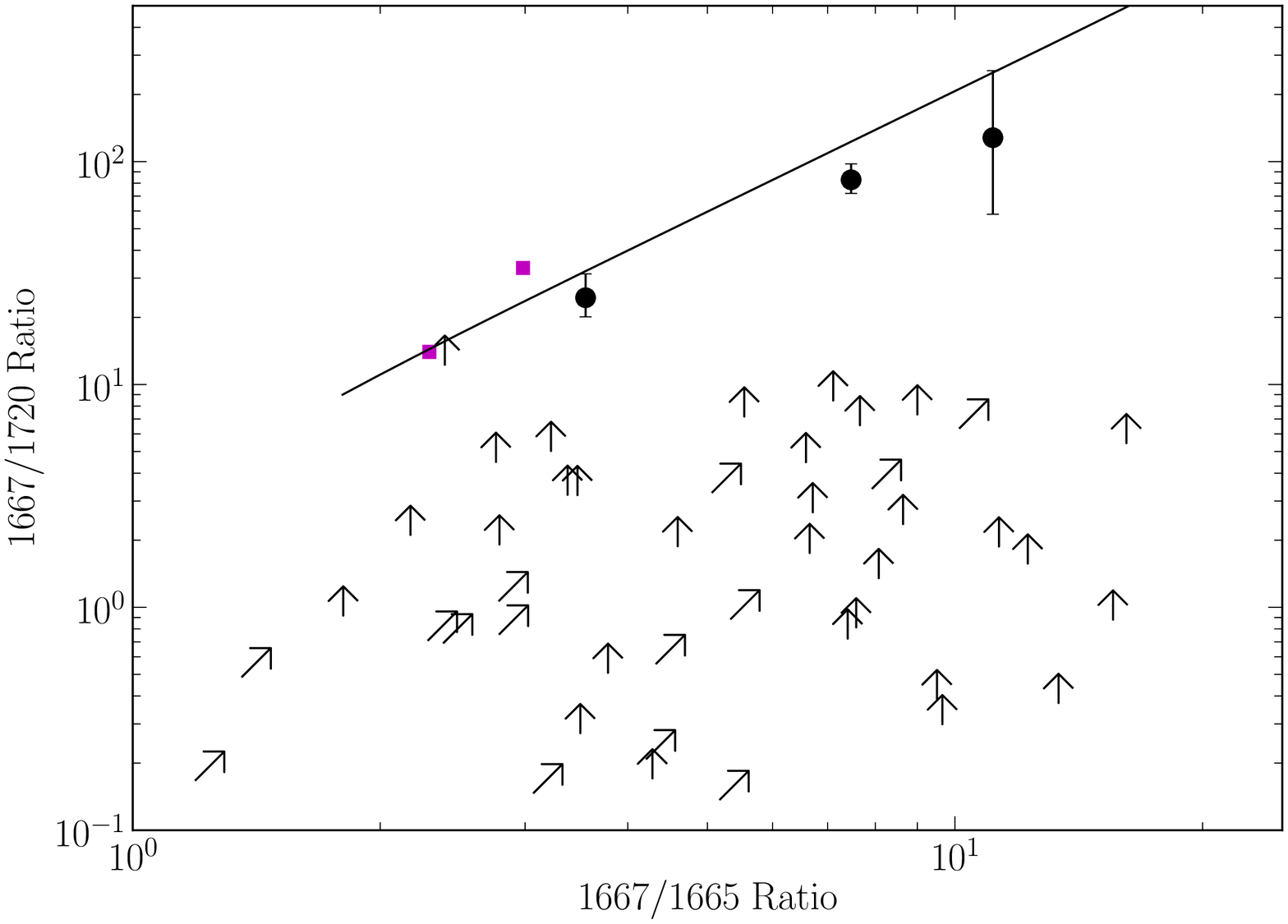}
    \label{fig:1720}
    }
    \subfigure[Excitation of the 1612~MHz line. The symbols are the same as in panel (a), but for the 1612~MHz line rather than the 1720~MHz line, and the line is given by Equations \ref{eq:excitation1665} and \ref{eq:excitation1612}.]{
    \includegraphics[width=3.5in] {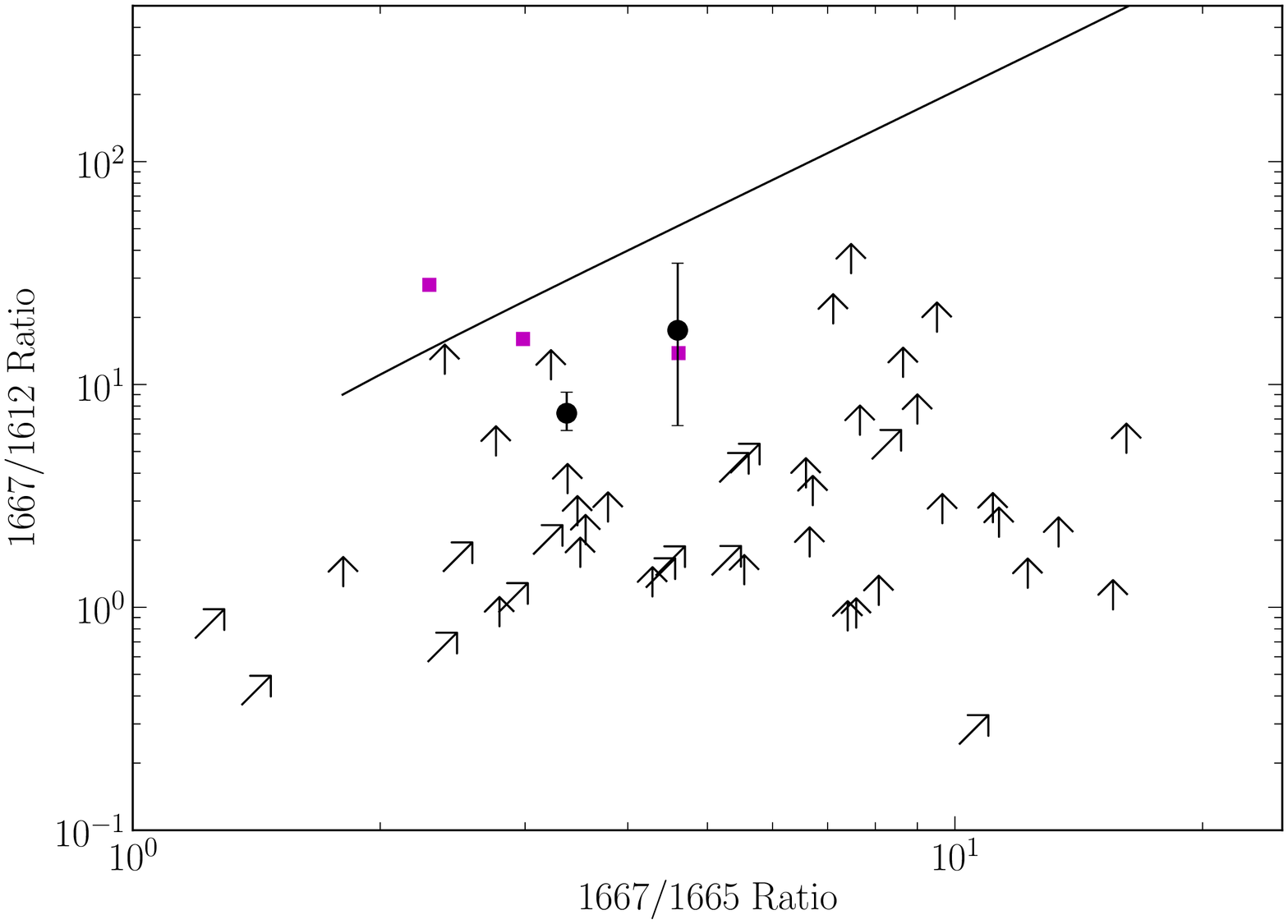}
    \label{fig:1612}
    }
\end{figure}

The explanation of OHM emission provided by HW90 assumed that the excitation temperatures of the OH 18~cm main lines are roughly the same. Multiple lines of evidence were provided for this assumption, including observations of other OH lines in the rotational ladder and earlier radiative transfer modeling by \citet{Henkel1987}. Equal excitation temperatures of the 18~cm OH lines was a result of the LE08 model calculations, which they noted occurred because of the line overlap produced by intrinsic linewidths of $\Delta V \geq 10\; \kms$. Observations of 18~cm OH satellite lines in OHMs provide an important test of equivalent excitation temperatures, as the lines should each be amplified according to their LTE line ratios. Thus, the observed line ratios should relate to the optical depth of the 1667~MHz line, $\tau$, as
\begin{align}
    \rh & = \frac{e^{-\tau} - 1}{e^{-\tau / 1.8} - 1}, \label{eq:excitation1665}\\ 
    \rtwelve & = \frac{e^{-\tau} - 1}{e^{-\tau / 9} - 1}, {\rm \; and} \label{eq:excitation1612} \\ 
    \rtwenty & = \frac{e^{-\tau} - 1}{e^{-\tau / 9} - 1}. \label{eq:excitation1720}
\end{align} 
LE08 plotted $\rh$ and $\rtwenty$ (in their Figure 6) for the 4 OHMs then known with detected 1665, 1667, and 1720~MHz emission, calling it a ``color-color'' diagram for OH maser lines. They found very nice agreement between the observed ratios and the expectation for equivalent excitation temperatures. We reproduce this plot, including all OHMs in the literature, our detections, and upper limits on the non-1667~MHz lines (meaning lower limits on the line ratios), for both the 1720~MHz line (Figure \ref{fig:1720}) and the 1612~MHz line (Figure \ref{fig:1612}).

The large number of non-detections makes a detailed test of equal line excitation impossible. Even so, some fraction of the lower limits on the ratios of 1667~MHz emission to satellite line emission are physically interesting. While IRAS~F15107+0724 has 1612~MHz emission that is considerably stronger than would be expected for equal line excitation temperatures, it is the only such example in the survey, and a few of the other sources have lower limits on the 1667/1612 ratio that preclude stronger-than-expected 1612~MHz emission of the nature seen in IRAS~F15107+0724. 

For the 1720~MHz line, the sources detected in emission all lie close to the line of equal excitation temperature, within error. IRAS~F15107+0724 is again the exception, as its 1720~MHz line appears in absorption, and for that reason is left out of Figure \ref{fig:1720} altogether. In the observed sample, there is no example of an OHM with 1720~MHz emission that is significantly stronger over the entire line profile than that which would be expected from equal excitation temperatures in the lines. The relative strength of the main lines and satellite lines {\em does} vary over different parts of the spectrum. For example, in IRAS~F10173+0829, the 1720~MHz emission is only visible in the region where 1667~MHz emission is weakest, and is absent over the region where 1667~MHz emission is strongest. \citet{Baan1987a} observed similar variation within the spectrum of Arp~220, and from that concluded that excitation conditions of the lines were different from one another within the same region. 

Overall, the relative weakness of satellite line emission as compared to 1667~MHz emission in OHMs is consistent with the results of HW90 and the modeling of LE08, in which all four of the 18~cm OH lines have roughly equal excitation temperatures. In sources with detected satellite lines, the lines are generally seen only within sub-regions of the 1667~MHz emission, and appear to be moderately stronger than expected for equivalent excitation temperatures. This suggests that secondary pumping mechanisms may occasionally contribute within OHMs, but 53~$\mu$m radiative pumping is the dominant pumping mechanism in the OHM population.

\section{Summary}
For the overwhelming majority of the 77 sources in this survey, no satellite line emission was detected. This result confirms that OHMs have
emission that is predominantly in the main lines, with most of the main
line emission at 1667~MHz. While this result is not unexpected, based
on the limited observations of satellite lines in OHMs prior to this survey and
on the best current models of emission in OHMs, it provides an important
additional constraint on the nature of OHMs. While some sub-regions of OHMs
display moderately stronger-than-expected satellite lines, relative to models 
in which there is roughly equal excitation of the 18~cm OH lines, there is 
not a large population of strong satellite line emitters among
the OHM population as a whole. This supports the results of 
LE08, who found that 53~$\mu$m radiative pumping, 
coupled with line overlap effects, dominates all other pumping
mechanisms in OHMs. The model assumptions in LE08 drew upon
the conclusions of \citet{Parra2005a}, who performed detailed modeling of the
emission in IRAS~F01417+1651, and concluded that the source could be well
explained by a clumpy ring of molecular gas in which masing clouds have
typical sizes of $\sim$1~pc, densities of $10^4\; \cmc$, and turbulent line 
widths of $\sim20\; \kms$. While our results do not directly test that
model, they are consistent with the general parameters they suggest.  

The OHM with the most prominent satellite
lines relative to its main lines is IRAS~F15107+0724, which is at the low
end of the OHM luminosity distribution. The observed properties of 
IRAS~F15107+0724 suggest a transition 
between OHMs such as Arp~220, in which all lines appear in emission and the
main lines are dominant, and OH kilomasers such as Messier~82 and NGC~253, 
in which narrow main line emission occurs within regions of absorption 
and satellite line features are more prominent. In the ``color-color''
diagrams of Figure \ref{fig:1720} and \ref{fig:1612}, the integrated
flux ratios of kilomasers would generally reside in the lower left corner,
away from the line showing expected flux ratios for equal excitation
temperatures of the lines. While there is significant variety within the
known kilomasers (HW90), they tend to have profiles dominated by absorption, 
hyperfine ratios that are approximately in the LTE range, and anomalies in
their satellite line strengths relative to LTE. This includes anomalously
strong or weak absorption in satellite lines, as seen in NGC~253
\citep{Gardner1975}, as well as satellite line emission accompanying main
line absorption, such as found in NGC~4945 \citep{Whiteoak1975}. Despite
the imhomogeneity in the kilomaser population, IRAS~F15107+0724 is still 
clearly apart from kilomasers in this respect, 
while also apparently unusual among the OHM population.

The overall low number of detections indicate that any future efforts to 
detect satellite lines in OHMs will require significantly better sensitivity 
than current radio telescopes can provide. 
The Five-hundred-meter Aperture Spherical Telescope (FAST), with an expected completion date in the next few years, could expand the sample of satellite
line detections in OHMs with a few tens of hours of observing time. 
Among sources with undetected satellite lines, IRAS~F02524+2046, IRAS~F09539+0857, IRAS~F15887+1609, IRAS~F23028+0725, and IRAS~F23129+2548 make the best targets for future observations with FAST, based on the brightness of their 1667~MHz
lines, hyperfine ratios, and OHM excitation models in which the 18~cm OH lines
have equivalent excitation temperatures. 
Detecting satellite lines in even lower flux OHMs in comparable amounts of 
time will require a more significant improvement in sensitivity, 
such as that which could be provided by the Square Kilometer Array.\\

We thank the anonymous referee for helpful suggestions, and Leo Blitz for useful discussions. This research was supported in part by NSF grant AST-0908572. J. M. received support from a National Science Foundation Graduate Research Fellowship. This research used NASA's Astrophysics Data System Bibliographic Services, the SIMBAD database, operated at CDS, Strasbourg, France, and the NASA/IPAC Extragalactic Database (NED), which is operated by the Jet Propulsion Laboratory, California Institute of Technology, under contract with the National Aeronautics and Space Administration.

\bibliographystyle{apj}
\bibliography{ms,extra}

\end{document}